\documentclass[11pt]{article}
\usepackage{moriond,epsfig,amssymb}

\def\Journal#1#2#3#4{{#1} {\bf #2}, #3 (#4)}

\def\EPJ{{\em Eur. Phys. J.} C}

\def\be{\begin{equation}}
\def\ee{\end{equation}}
\def\bea{\begin{eqnarray}}
\def\eea{\end{eqnarray}}

\begin{document}
\vspace*{4cm}
\title{CHARM IN ELECTRON--PROTON COLLISIONS}

\author{ A. BERTOLIN \\
on behalf of the ZEUS and H1 collaborations}

\address{Istituto Nazionale di Fisica Nucleare,
Sezione di Padova \\
Via Marzolo 8, 35131 Padova, Italy}

\maketitle\abstracts{
Charm in electron--proton collisions is a particularly large and
interesting field. For this summary two topics particularly relevant
for HERA have been selected.
First the study of inelastic $J/\psi$ production in both the 
photoproduction and the electroproduction regime, study still needed to 
clarify the $J/\psi$ production mechanisms.
Second the analysis of $D^{\star}$ photoproduction events, study than
can lead to a better understanding of the photon structure.}

\section{Inelastic $J/\psi$ photoproduction}

\begin{figure}[hbpt!]
\hspace{0.5cm} \psfig{figure=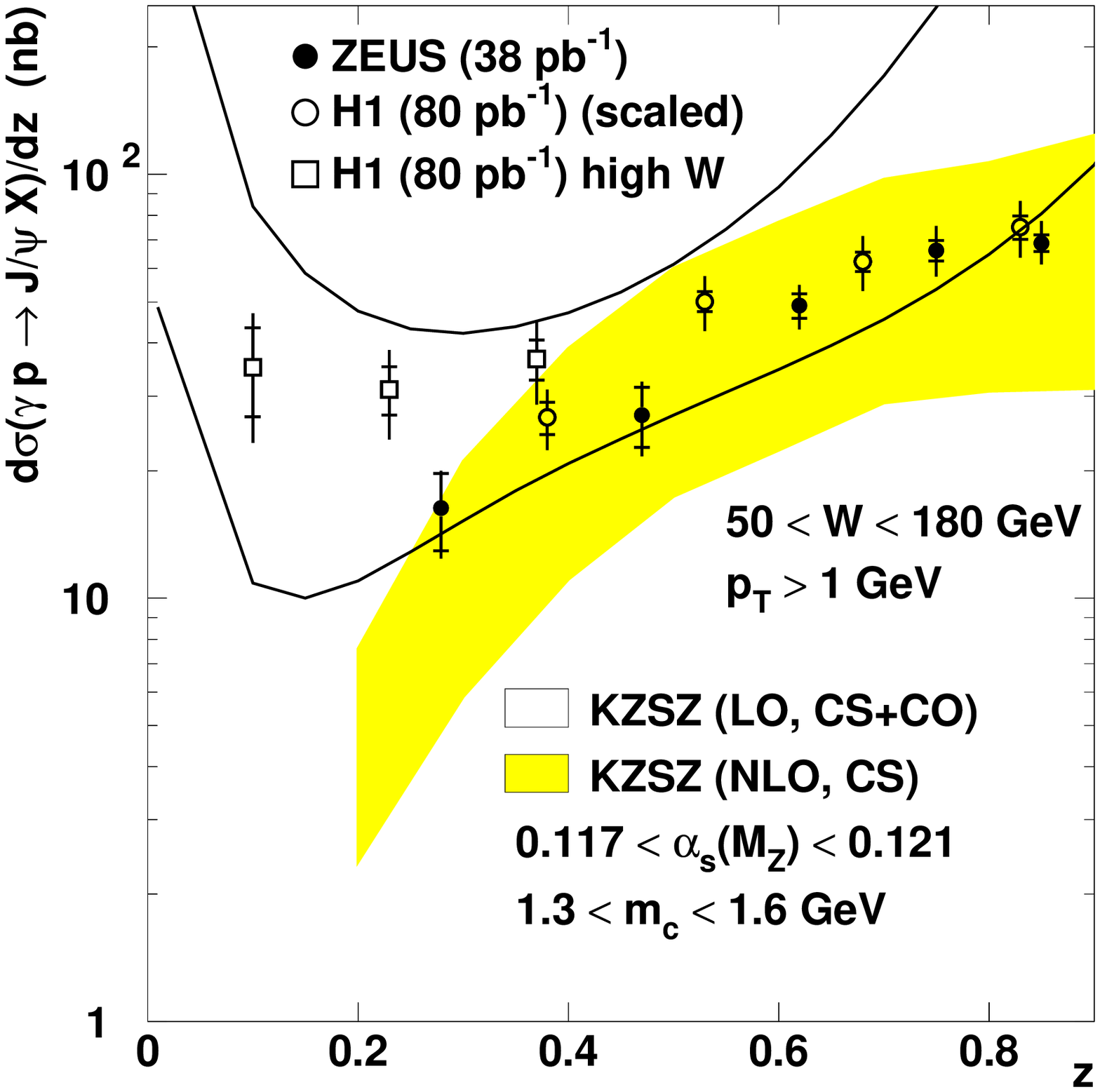,height=2.75in}
\hspace{0.5cm} \psfig{figure=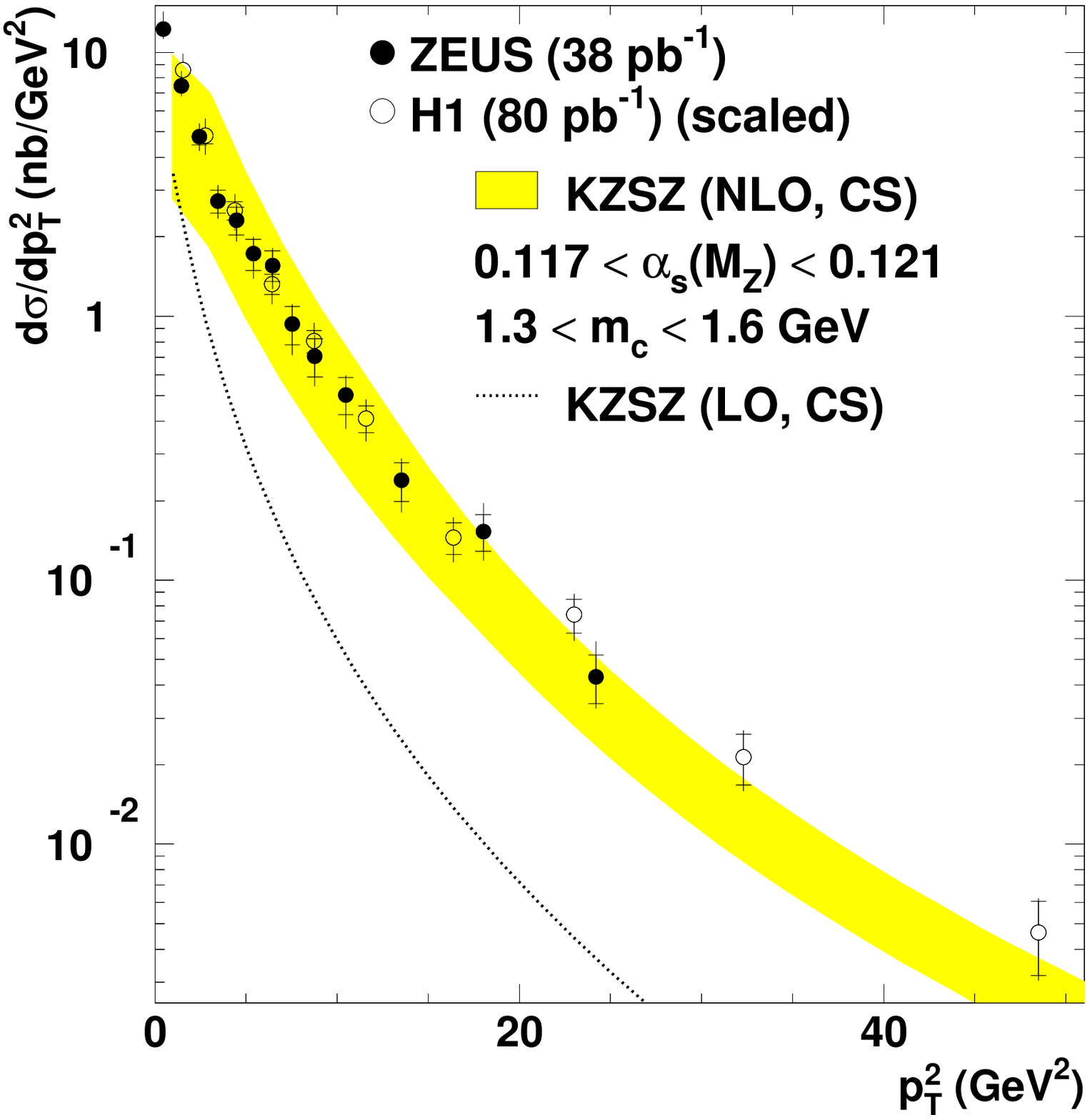,height=2.75in}\\
\psfig{figure=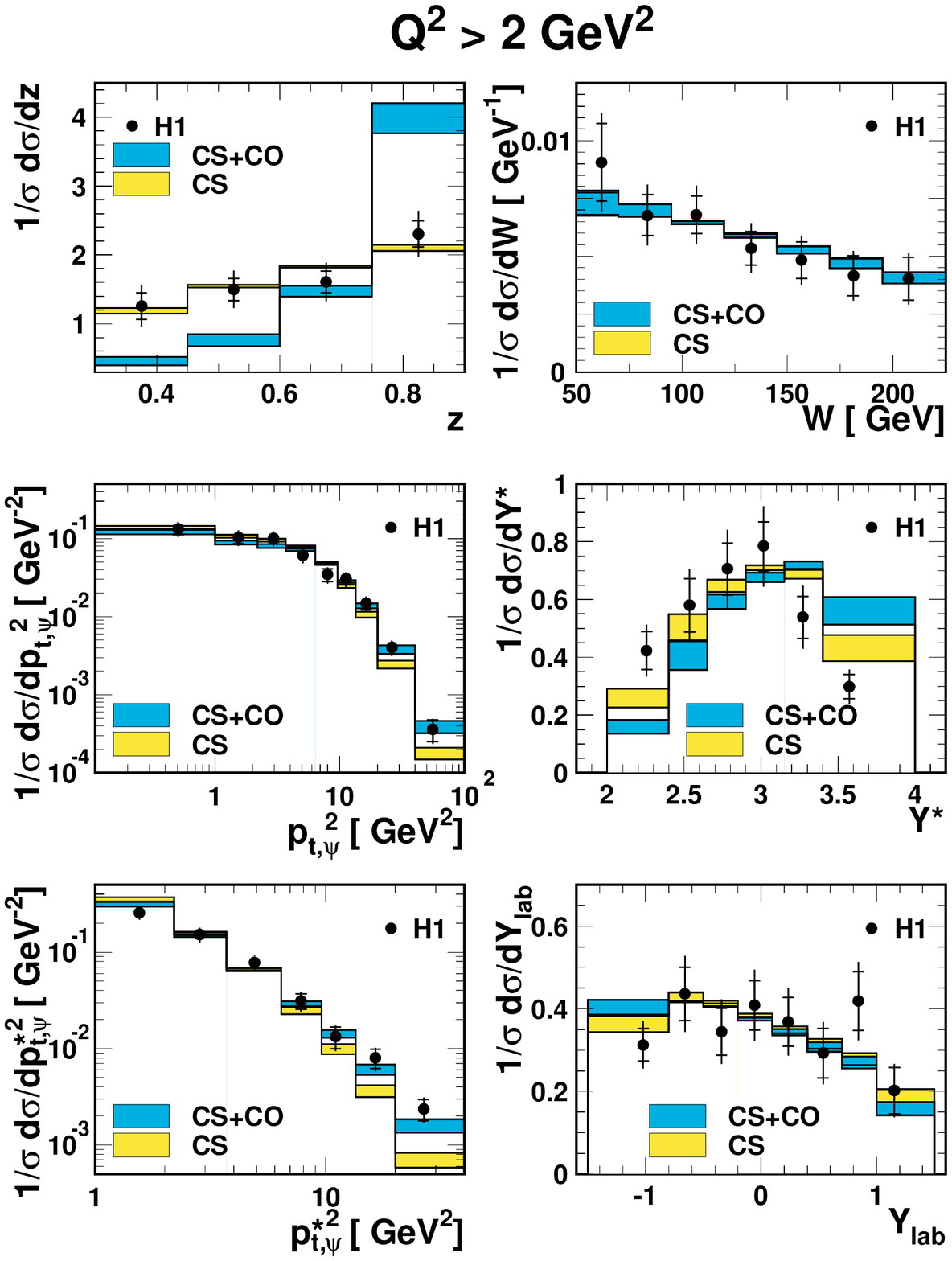,height=3.75in}
\label{ij}
\caption{Inelasticity, $z$, upper left, and $p_T^2$, upper right, 
differential cross section measured by the ZEUS and H1 collaborations 
for the inelastic $J/\psi$ photoproduction process. The theoretical 
predictions are described in the text.
The bottom left plot shows the inelasticity, $z$, photon proton
center of mass energy, $W$, $J/\psi$ transverse momentum squared,
$p_{t,\psi}^2$, $J/\psi$ rapidity in the $\gamma^{\star}$ $p$ frame,
$Y^{\star}$, $J/\psi$ transverse momentum squared in the $\gamma^{\star}$ 
$p$ frame, $p_{t,\psi}^{\star 2}$ and $J/\psi$ rapidity in the lab.
frame, $Y_{lab}$, differential cross sections measured by the H1 
collaboration for the inelastic $J/\psi$ electroproduction process. 
The theoretical predictions are described in the text.}
\end{figure}

The inelastic $J/\psi$ photoproduction process has been measured by
both the H1~\cite{h1ijphp} and ZEUS~\cite{zijphp} collaborations. The
key variable describing this process is the inelasticity, $z$. In the
proton rest frame $z$ is the fraction of the virtual photon energy taken
by the $J/\psi$. Resolved photon processes are expected to dominate at low 
$z$ while for $0.2 \le z \le 0.9$ direct photon processes take over
and hence the characteristic inelasticity shape shown in the top left
plot of fig. \ref{ij} by the LO calculation labelled KZSZ (LO,CS+CO).
The production at higher $z$ values is ascribed to diffractive processes.
Two different categories of subprocesses contribute to the direct and to 
the resolved part, called colour singlet, CS, and colour octet, CO, 
processes. The introduction of the CO component is a consequence of non 
relativistic QCD theories and is needed to explain the magnitude of the
$J/\psi$ cross section seen by CDF.
The HERA data shown in the top left plot of fig. \ref{ij} are 
consistent with these prediction although the rise of the inelasticity 
cross section for $z \gtrsim 0.6$ is softer in the data with respect to 
the KZSZ (LO,CS+CO) prediction.
The H1 data for $z \gtrsim 0.4$ have been scaled to take into account a 
small difference in the $W$ integration range between the ZEUS and the H1
analyses.
The low $z$ H1 data, $z \lesssim 0.4$, are integrated for $120 < W < 260$
GeV. The different $W$ integration range is not taken into
in the data or the theoretical predictions.
The measured inelasticity differential cross section is also compared to 
a NLO calculation including only the direct photon colour singlet process, 
KZSZ (NLO,CS).
This calculation, available only for $0.2 < z < 0.9$, describes, within 
large theoretical uncertainties, both the shape and the normalization 
of the data.
The same calculation can also account for the $p_T^2$ differential cross
section, as shown by the top right plot of fig. \ref{ij}. In the HERA 
case the NLO corrections to the direct photon colour singlet process are 
found to be large and increasing with $p_T^2$, as can be seen by comparing 
the LO and NLO predictions.

\section{Inelastic $J/\psi$ electroproduction}

The inelastic $J/\psi$ process in the high $Q^2$ regime, the 
electroproduction regime, has been measured by the H1 
collaboration~\cite{h1ijdis}: the inelasticity, $z$, photon proton
center of mass energy, $W$, $J/\psi$ transverse momentum squared,
$p_{t,\psi}^2$, $J/\psi$ rapidity in the $\gamma^{\star}$ $p$ frame,
$Y^{\star}$, $J/\psi$ transverse momentum squared in the $\gamma^{\star}$ 
$p$ frame, $p_{t,\psi}^{\star 2}$ and $J/\psi$ rapidity in the lab.
frame, $Y_{lab}$, are shown in bottom left series of plots in 
fig. \ref{ij}~\footnote{ZEUS preliminary results on inelastic $J/\psi$
electroproduction have been shown at the 11th International Workshop on Deep 
Inelastic Scattering, 23-27 April 2003, St. Petersburg, Russia.}. 
The data are compared to a LO calculation including the CS component only, 
identified by the label CS, or including both the CS and the CO component, 
CS+CO.
To avoid the large normalization uncertainties of a LO calculation
only the shapes of the various distributions are considered. 
As in the photoproduction case, the key variable is the inelasticity, $z$:
the shape measured in the data clearly favor the CS component only 
hypothesis.

\section{$D^{\star}$ photoproduction}

$D^{\star}$ photoproduction has been investigated in detail by the
ZEUS collaboration~\cite{zdstarphp}. Measurements of the $D^{\star}$
transverse momentum, $p_T(D^{\star})$, $D^{\star}$ pseudorapidity,
$\eta(D^{\star})$, photon proton center of mass energy, $W$, and
$D^{\star}$ inelasticity, $z$, differential cross sections are shown
in the left part of fig. 2. 
The data are compared to a NLO QCD calculation, identified by the label 
NLO QCD, and to a calculation incorporating mass effects up to the NLO 
and resuming $p_T$ logarithms up to the NLL level, labelled FONLL, 
available only for fig. 2 (a) and (b).
The NLO QCD calculation can describe the data in the forward region
and at low $z$ values only using extreme values for the non perturbative
input parameters ($m_c =$ 1.3 GeV). Instead for $0.6 < z < 1$, see fig.
2 (d), the normalization of the NLO QCD prediction
matches the data. It can be shown that this region corresponds mostly
to the direct photon process. \\
In $D^{\star}$ events with two hadronic jets the separation between direct 
and resolved photon processes can also be achieved using the variable 
$x_{\gamma}^{obs}$ defined as:
\begin{center}
$x_{\gamma}^{obs} = \frac{\Sigma_{jets} E_t^{jet} e^{-\eta^{jet}}}
                         {2 y E_e}$
\end{center}
where $E_t^{jet}$ and $\eta^{jet}$ are the transverse energies and the
pseudorapidities of the two jets, respectively, $E_e$ is the electron
beam energy and $y$, in the proton rest frame, is the fraction of the
electron energy take by the quasi--real exchanged photon.
Direct and resolved photon events are selected by requiring $x_{\gamma}^{obs}$
to be above or below the threshold value of 0.75, respectively.
The right part of fig. 2 shows the $\cos \theta^{\star}$ 
differential cross section, where $\theta^{\star}$ is the angle between the 
jet--jet axis and the proton--photon axis in the parton--parton rest frame, 
for direct and resolved photon events. This distribution should behave
like:
\begin{center}
$d\sigma / d |\cos \theta^{\star}| \approx 
(1-|\cos \theta^{\star}|)^{-2 S}$
\end{center}
where $S$ is the spin of the exchanged propagator.

The data for $x_{\gamma}^{obs} < 0.75$, the resolved photon region, 
show a steeper rise of the cross section in the photon direction
with respect to the $x_{\gamma}^{obs} > 0.75$ selection, the direct
region. Hence gluon exchange is the dominant mechanism in the
$x_{\gamma}^{obs} < 0.75$ region.
Furthermore since the charm jet is close to the photon in rapidity
charm must come from the photon.
This class of diagrams, called charm flavour 
excitation in the photon, are implemented in the PYTHIA MC which 
describes the $\cos \theta^{\star}$ differential cross section for 
both $x_{\gamma}^{obs}$ selections.
The HERWIG MC gives also a good description of the data for 
$x_{\gamma}^{obs} > 0.75$ but is worse than PYTHIA for 
$x_{\gamma}^{obs} < 0.75$.
In the CASCADE MC the $k_T$ unintegrated gluon density of the proton 
obey the CCFM evolution equation. In this case even if the CASCADE matrix 
elements correspond to the direct photon process only resolved photon 
processes are reproduced by the CCFM initial state radiation.
CASCADE fails to reproduce the data for $x_{\gamma}^{obs} > 0.75$
but gives a good description of the data for the lower $x_{\gamma}^{obs}$
range.
A NLO QCD calculation, including hadronization corrections, identified
by the label NLO QCD $\otimes$ HAD, gives a good description of the
data for $x_{\gamma}^{obs} > 0.75$ but is only marginally consistent 
with the data for $x_{\gamma}^{obs} < 0.75$ in the photon direction.

\begin{figure}[hbpt!]
\hspace{-1cm} \psfig{figure=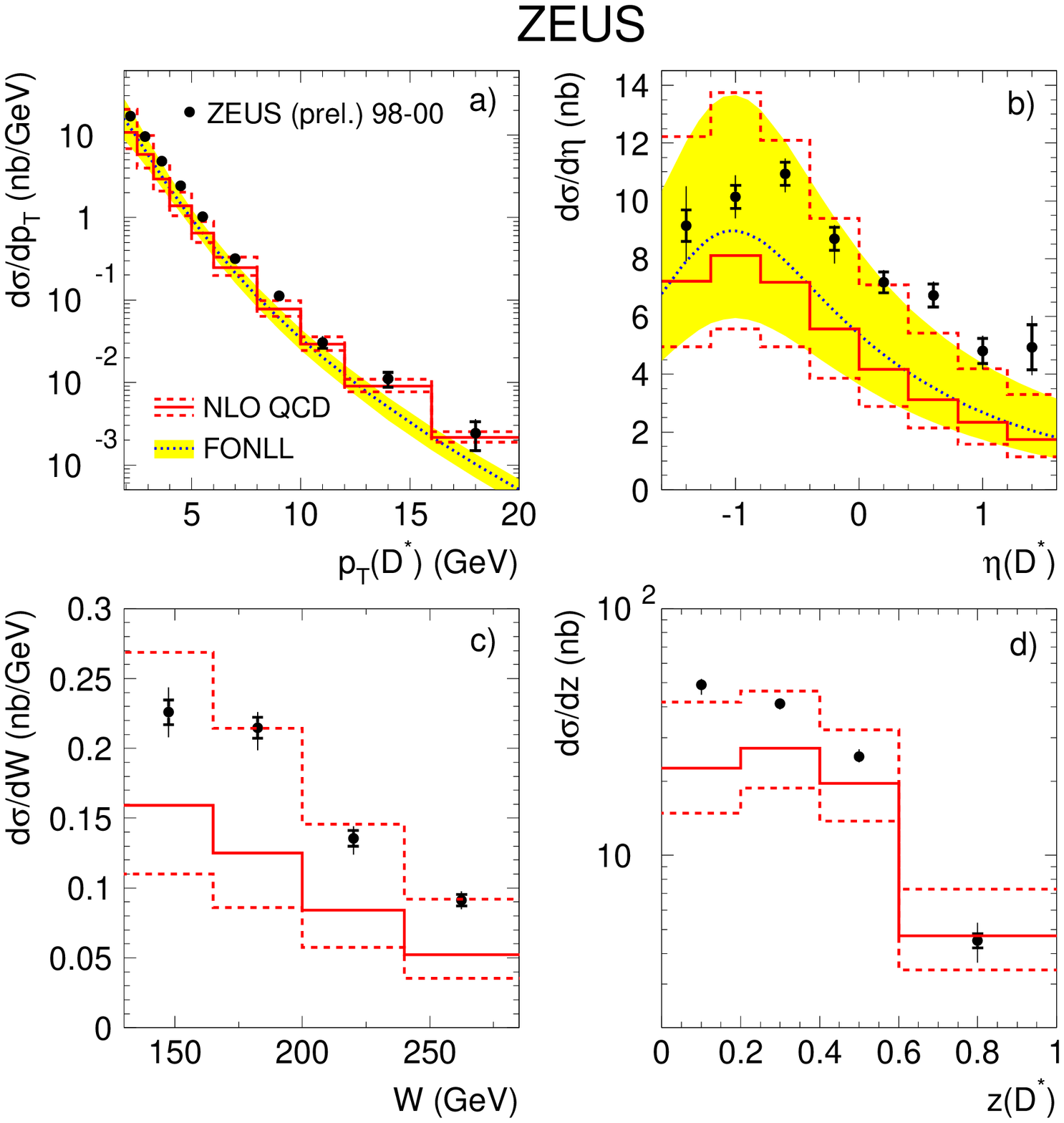,height=4.25in}
\hspace{-1.25cm} \psfig{figure=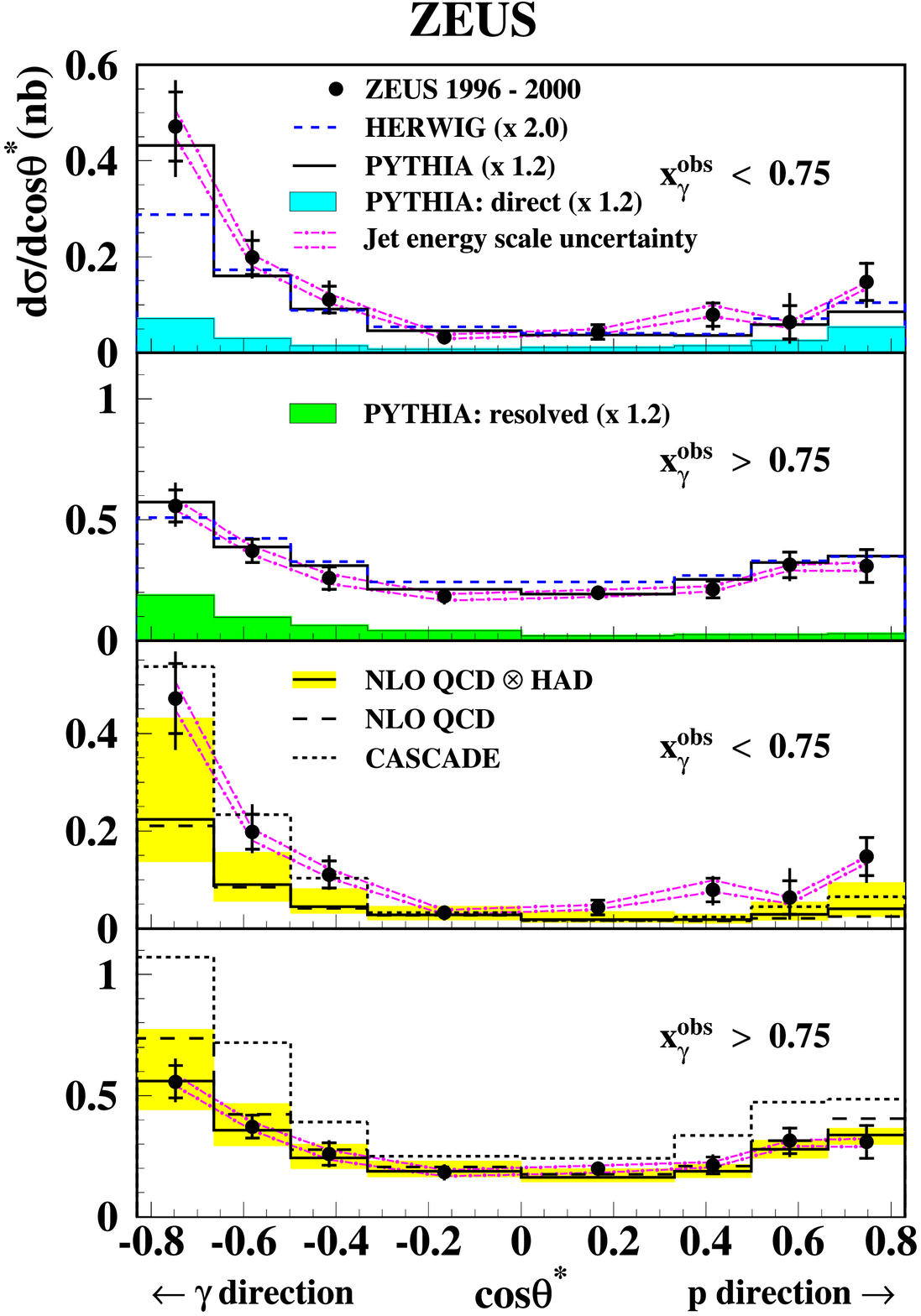,height=4.25in}
\label{dsphp}
\caption{Measurements of the $D^{\star}$ transverse momentum, 
$p_T(D^{\star})$, $D^{\star}$ pseudorapidity, $\eta(D^{\star})$, photon 
proton center of mass energy, $W$, and $D^{\star}$ inelasticity, $z$, 
differential cross sections are shown in fig. from (a) to (d), respectively.
The right side compares the $\cos \theta^{\star}$ differential cross section,
where $\theta^{\star}$ is the angle between the jet--jet axis and the 
proton--photon axis in the parton--parton rest frame, for two different
$x_{\gamma}^{obs}$ selections, to MC models and QCD predictions.
All models and predictions are described in the text.}
\end{figure}




\section*{References}
\begin{thebibliography}{99}

\bibitem{h1ijphp} C. Adloff {\it et al}, \Journal{\EPJ}{25}{25}{2002}.

\bibitem{zijphp} S. Chekanov {\it et al}, \Journal{\EPJ}{27}{173}{2002}.

\bibitem{h1ijdis} C. Adloff {\it et al}, \Journal{\EPJ}{25}{41}{2002}.

\bibitem{zdstarphp} S. Chekanov {\it et al}, DESY-03-015, submitted to 
{\em Phys. Lett.} B; \\
S. Chekanov {\it et al}, Abstract 786 Submitted to the XXXIst International
Conference on High Energy Physics, 24-31 July 2002, Amsterdam,
The Netherlands.





\end{thebibliography}
\end{document}